\documentstyle[12pt,epsfig]{article}

\def \lsim{\mathrel{\vcenter
     {\hbox{$<$}\nointerlineskip\hbox{$\sim$}}}}
\def \gsim{\mathrel{\vcenter
     {\hbox{$>$}\nointerlineskip\hbox{$\sim$}}}}
\def\beq{\begin{equation}}
\def\eeq#1{\label{#1}\end{equation}}
\def\eeqn{\end{equation}}

\def\beqa{\begin{eqnarray}}
\def\eeqa#1{\label{#1}\end{eqnarray}}
\def\eeqan{\end{eqnarray}}
\def\CR{\nonumber \\ }

\def\leqn#1{(\ref{#1})}

\let\bar=\overbar

\def\Dslash{\not{\hbox{\kern-4pt $D$}}}
\def\dslash{\not{\hbox{\kern-2pt $\del$}}}

\def\msb{{\bar{\ssstyle M \kern -1pt S}}}

\def\Title#1{\begin{center} {\Large {\bf #1} } \end{center}}

\begin{document}

\Title{Recent Developments in Physics Beyond the Standard 
Model\footnote{Plenary talk at the XIX International Symposium on Lepton and
Photon Interactions at High Energies, Stanford University, 9-14 August 1999.}}

\bigskip\bigskip


\begin{raggedright}  

{\it G.F. Giudice\index{Giudice, G.F.}\\
Theory Division \\
CERN,
CH-1211 Geneva 23, Switzerland}
\bigskip\bigskip
\end{raggedright}

In this talk I discuss what I believe are the most interesting recent
developments in physics
beyond the Standard Model. After some initial comments on neutrino masses,
I discuss the status of low-energy supersymmetry and finally turn to
describing some recent work in theories with extra spatial dimensions.

\section{Neutrinos}

The most concrete indication for the existence of physics beyond the Standard
Model has recently emerged from the Superkamiokande data~\cite{superk}, which
convincingly confirm the presence of an 
atmospheric neutrino anomaly~\cite{atmos}. The most reasonable explanation
for these experimental observations relies on the assumption
that neutrinos are massive and that the different flavour eigenstates can 
oscillate among each other. If this interpretation is correct, it implies
evidence for new physics beyond the Standard Model.
During this Conference, we have
heard much discussion of the experimental status of neutrino masses and 
oscillations. Let me here make some comments on what I believe are the
main lessons for theory 
we have learnt from these results which, if confirmed, represent
one of the most important discoveries in physics in recent years. 

\bigskip
$\bullet$ We are finding evidence for
a new mass scale much larger than the typical weak scale, but
different from the Planck mass $G_N^{-1/2}$.
Indeed, including only
Standard Model degrees of freedom, neutrino masses are described by
dimension-5 operators of the form
\beq
\frac{1}{\Lambda}~ {\ell}_L^T C \ell_L~HH,
\eeq{op}
where $\ell_L$ represents the lepton weak doublet and $C$ is the 
charge-conjugation matrix. After electroweak symmetry breaking, the Higgs
field $H$ gets a non-vanishing vacuum expectation value, 
and the operator in eq.~\leqn{op}
leads to a Majorana neutrino mass $m_\nu =\langle H\rangle^2/\Lambda$.
According to the most natural interpretation of the atmospheric neutrino
anomaly, there exists a neutrino with mass of about $6\times 10^{-2}$~eV.
This implies a new-physics mass scale $\Lambda$ at about $10^{15}$~GeV,
tantalizingly close to the GUT scale. 

\bigskip
$\bullet$ We are finding that one neutrino mixing angle (most likely the
one corresponding to $\nu_\mu$--$\nu_\tau$ oscillations) is large, since
the best fit of the Superkamiokande data gives $\sin^2 2\theta =0.99$
and $\Delta m_\nu^2=3.1\times 10^{-3}$~eV$^2$~\cite{tmann}. This
situation is different from the case of the familiar 
Cabibbo--Kobayashi--Maskawa (CKM) mixing among quarks, and therefore it 
first appeared
as a surprise. To assess if this result contradicts our prejudices on the
structure of Yukawa couplings, we have to understand if it is indeed 
incompatible with hierarchical neutrino masses and with GUTs, which relate
the properties of quarks and leptons. Recent investigations~\cite{neutr}
have shown that this is not the case; let me explain why.

Let us consider the neutrino mass matrix as it emerges from the see-saw
mechanism:
\beq
m_\nu =h_\nu^T ~M^{-1}~ h_\nu \langle H \rangle^2.
\eeq{mnu}
Here $h_\nu$ is the $3\times 3$ Yukawa coupling matrix and $M$ is the 
right-handed neutrino Majorana mass matrix. If the large mixings reside
in the matrix $M$ but not in $h_\nu$, the neutrino oscillation results can
be simply made compatible with $SU(5)$ GUTs relations, since the right-handed
neutrinos are $SU(5)$ singlets. An interesting 
alternative~\cite{alta}
is that the large mixings reside instead in the left-handed neutrinos. This
is compatible with $SU(5)$ if the Yukawa coupling matrices are highly
asymmetric. Indeed, the $SU(5)$ relation between the charged lepton and
down quark Yukawa coupling is $h_\ell =h_d^T$. Therefore a large mixing
in the left-handed
charged lepton sector (which corresponds to large neutrino mixing 
after an $SU(2)$ rotation) corresponds to a large mixing in the right-handed
quark sector (which does not affect the CKM matrix). 

It has also been observed~\cite{alta2} that large neutrino mixing angles are
not incompatible with hierarchical structures in $h_\nu$ and $M$. For
instance, consider the toy model of $2\times 2$ symmetric matrices
\beq
h_\nu =\pmatrix{A\epsilon & B\epsilon \cr B\epsilon & 1},~~~~
M=\pmatrix{C\epsilon^n & D\epsilon^m \cr D\epsilon^m & 1},
\eeq{cazzinsu}
with $A,B,C,D$ parameters of order unity and $\epsilon \ll 1$. 
From eq.~\leqn{cazzinsu}, we find 
that, for $n>2$ and $m>n/2$, the two eigenvalues of the matrix  $m_\nu$
are $\epsilon^{2-n}(A^2+B^2)/C$ and $\epsilon^n A^2C/(A^2+B^2)$,
and therefore there is
a hierarchy between the two physical neutrino masses. On the other hand,
the neutrino mixing parameter $\sin 2\theta =2AB/(A^2+B^2)$ is of
order unity, as long as $A\simeq B$.

In conclusion, although it could not have been theoretically anticipated,
the large mixing angle suggested by atmospheric neutrino data can be easily
accommodated
both with hierarchical neutrino masses and with GUT relations.  

\bigskip
$\bullet$ It is well known that neutrino masses have profound consequences
in cosmology and astrophysics. I just want to emphasize here that the results
of the atmospheric neutrino data strongly suggest a GUT-realized see-saw
mechanism and therefore give further justification for 
leptogenesis~\cite{lepto}. Indeed,
I find that at present the best motivated way of explaining the observed
baryon asymmetry is to invoke the out-of-equilibrium decay of the right
neutrinos. With the natural assumption of the presence of CP-violating
phases in the Yukawa couplings, the right-handed neutrino decay modes
\beqa
N_R &\to & \ell_L H ,\\
N_R &\to & {\bar \ell}_L H^*
\eeqa{nrd}
give rise to a cosmic lepton asymmetry. Sphaleron-like interactions, which 
violate a certain linear combination of lepton and baryon number,
are
in thermal equilibrium before the electroweak phase transition, and reshuffle
the particle populations, creating a baryon asymmetry. It is very
encouraging that a large range of reasonable neutrino mass parameters can
lead to the correct value of the baryon asymmetry. The leptogenesis can then
also be used as a criterion to select or disfavour particular models
of fermion mass matrices. However, it is unfortunately hard to translate
the conditions for successful leptogenesis into simple constraints on the
observed neutrino masses and mixings. The main reason for this
is that leptogenesis is driven by the inclusive
decay processes (4) and (5), 
summed over the three generations of $\ell_L$. 
Therefore leptogenesis is mainly
sensitive to the mixing angles in the right-handed sector, while
experiments observe the properties of mainly left-handed
neutrinos.

\bigskip
$\bullet$ The ultimate goal of the theoretical activity is to use the
experimental information on neutrino masses and mixing in order to unravel 
the flavour mistery and construct a predictive theory of fermion masses.
Although there has been quite an intense research with this aim~\cite{neutr},
I believe that we are still far, unfortunately, from understanding the rationale
of the flavour structure.   

\section{Supersymmetry}

As we have discussed above, one of the most important consequence of
the atmospheric neutrino oscillations 
is the evidence for a new mass scale $\Lambda$.
In this respect, this result
agrees with the other indirect
indication for new physics: the unification of gauge coupling constants. They
are both hints to the existence of a physical threshold at the GUT scale.
Following this line of reasoning, one is almost compelled to believe in
low-energy supersymmetry. This is because a simple extrapolation of the
Standard Model to energies much larger than the weak scale requires a
disturbing fine tuning of the parameters in the Higgs potential, while
supersymmetry allows for such an extrapolation without conflicts with 
criteria of naturalness. Moreover, the prediction of 
$\alpha_s$ under unification assumptions fails if the $\beta$ functions
contain only the contributions from Standard Model particles, but it
correctly reproduces the experimental value when one includes the quantum
effects of the supersymmetric partners with masses in the 100 GeV--1 TeV range.
Therefore, the theoretical motivations for low-energy supersymmetry are
still very strong.

On the other hand, the experimental limits~\cite{franc} 
are worryingly increasing. The
limit on the chargino mass is 100 GeV (except for certain pathological regions
of parameter space), the one on the lightest
neutralino mass  is 37 GeV (assuming
GUT-related gaugino masses). In the minimal scheme, the gluino mass 
limit varies between about 200 GeV (for very large squark mass ${\tilde m}_q$)
to 300 GeV (for ${\tilde m}_g \simeq {\tilde m}_q$). A considerable constraint
also comes from the Higgs mass limit, which varies between 90 GeV (for large
$\tan
\beta$) to
106 GeV (for small
$\tan \beta$). It seems appropriate and timely to question whether these
limits are compatible with the original motivation for low-energy supersymmetry,
{\it i.e.} the hierarchy problem. 

To obtain a quantitative answer, one has to rely on a naturalness 
criterion~\cite{natural1,natural2,natural3} 
which specifies the amount of fine tuning among parameters.
Recent analyses quantify the result in different ways and conclude
that the present experimental limits rule out ``95\% of the supersymmetric
parameter space"~\cite{romans} or require ``fine tunings among parameters
at a level of 7\% or more"~\cite{pol}. Undoubtedly these statements sound
rather grim. However it should be noted that they are based on certain
theoretical assumptions and prejudices. For instance, for specific values
of the top quark Yukawa coupling (corresponding to not too small values
of $\tan \beta$) a universal squark and slepton mass contribution at the
GUT scale cancels out in the expression of $M_Z$~\cite{natural2,feng}. 
This means that, in this
case, squarks and sleptons can be made heavy without causing serious
fine-tuning difficulties. Depending on your favourite point of view,
this situation
can or cannot be viewed as an indirect fine tuning on the top Yukawa
coupling. Another interesting observation~\cite{wri} is that the naturalness
bounds on charginos and neutralinos are significantly modified if we 
abandon GUT relations on gaugino masses. This is because in the expression
of $M_Z$ in terms of the supersymmetry-breaking parameters, there is 
a prominent sensitivity on the gluino mass, but
only
a mild dependence on the electroweak gaugino masses. Fine tunings of no
more than 10\% can be achieved for chargino masses as large as 165 GeV, although
the gluino has to be lighter than 260 GeV. In conclusion, although the
present experimental bounds have severely limited the plausible range of
supersymmetric parameters, low-energy supersymmetry is far from being
ruled out and we have to wait for the LHC for the final verdict. 

Let us now turn to discussing the theoretical developments in supersymmetric
model building. Most of the recent activity has focused on understanding
the structure of the soft supersymmetry-breaking terms, especially in view
of the flavour problem I will illustrate below. The question of the origin
of the supersymmetry-breaking terms is indeed a crucial one, because the
soft terms represent the connection between theory ({\it i.e.} the mechanism
of supersymmetry breaking) and experiment ({\it i.e.} the mass spectrum of
the new particles).

For many years the paradigm has been that the soft terms are produced
by the gravitational couplings between a hidden sector where supersymmetry
is originally broken and an observable sector containing the ordinary
degrees of freedom~\cite{grav}. 
The scale of supersymmetry breaking is determined
to lie at an intermediate scale $\sqrt{F}\sim 10^{11}$~GeV by requiring
that the observable supersymmetric particle masses ${\tilde m}$
are close to the 
weak scale:
\beq
{\tilde m} =\frac{F}{M_{Pl}}\sim {\rm TeV}.
\eeqn

This mechanism is elegant and theoretically appealing, as gravity is directly
participating in electroweak physics. However, in this framework, the soft
terms are renormalizable parameters of the effective theory defined at
energies below the Planck mass $M_{Pl}$. As such, at the quantum level, 
they receive corrections that depend on the properties of the underlying
theory in the far ultraviolet. Therefore the soft terms cannot be computed,
as long as we do not know the ultimate theory including a full description
of quantum gravity. This could just be a limitation due to our lack of
knowledge but, from a pragmatic point of view, it introduces two main
problems. The first one is the lack of theoretical predictivity. This is
indeed an acute problem since, even with the minimal field content, the
low-energy supersymmetric model contains more than 110 free remormalizable
parameters, crippling our ability to give solid guidelines to experimental
searches. Secondly, the sensitivity of the soft terms to ultraviolet physics
implies that their flavour structure will retain the effects of any (unknown)
flavour violation at very high energies~\cite{kos}. In particular, flavour
universality of the soft
terms will be spoilt by new interactions, which include, for instance, effects
from GUTs~\cite{hall} or from the dynamics at the (unknown) scale $\Lambda_F$
responsible for the origin of the flavour-violating Yukawa couplings. This
is described by the lines indicated with the caption
``Gravity mediation" in fig.~1. This figure schematically illustrates
the energy dependence
of the running squark masses of the three different generations. Even if
we hypothetically took mass-degenerate squarks  at $M_{Pl}$, high-energy
flavour violations would induce large squark splittings, not correlated to
the Yukawa couplings, at low energy. This situation is experimentally
ruled out because the flavour violations in squarks and sleptons induce,
through loop diagrams, unacceptably large contributions to $\Delta m_K$,
$\epsilon_K$, $\Delta m_B$, $b\to s \gamma$, $\mu\to e \gamma$, and other
flavour processes.

\begin{figure}
\begin{center}
\epsfig{file=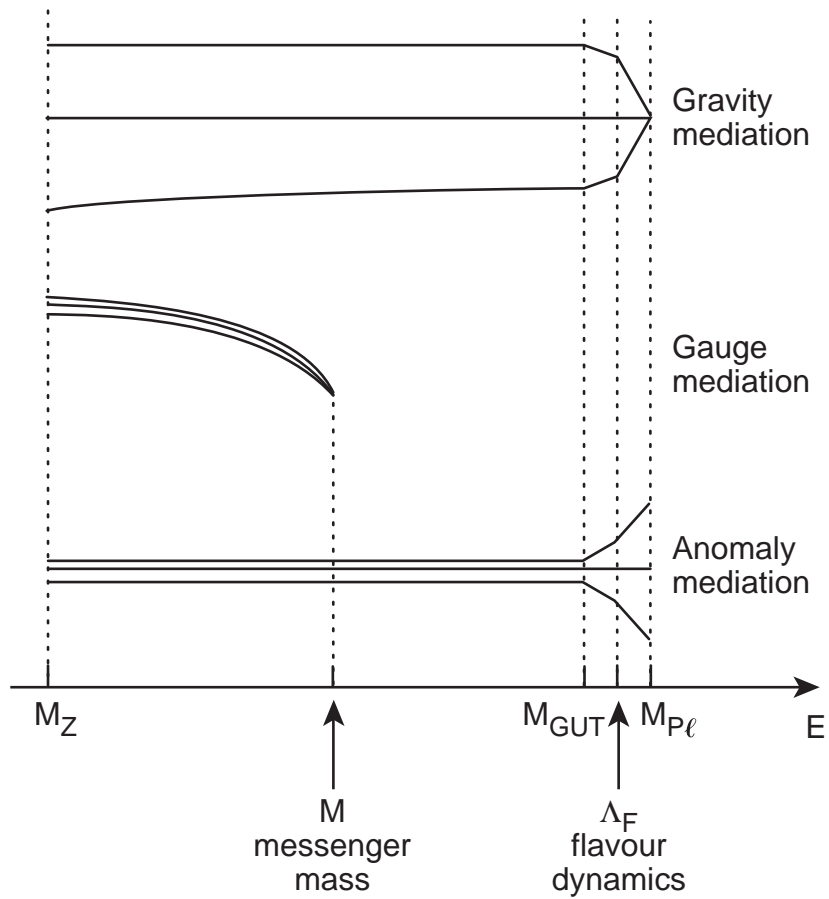}
\caption{A schematic illustration of the energy dependence of the running
squark masses belonging to the three different generations, in the context
of the various supersymmetric scenarios discussed in the text. In gravity
mediation, new dynamics at the scale $\Lambda_F$ and GUT physics tend to
induce large flavour-breaking effects in the squark spectrum, even if we
start from a universality assumption at $M_{Pl}$. In the case of gauge
mediation, the squark masses can be generated at scales sufficiently low
to ensure a super-GIM mechanism. In anomaly mediation, the squark spectrum 
is determined by the low-energy theory and it is insensitive to 
flavour violations occurring at large scales.}
\end{center}
\end{figure}

To solve the flavour problem in the context of gravity-mediated supersymmetry
breaking one needs to have full control of the dynamics even beyond 
$M_{Pl}$. It is possible that its solution lies in the properties of
quantum gravity and its flavour symmetries. However, recently there have been 
theoretical attempts to pursue alternative solutions, finding mechanisms
aimed at eliminating 
the ultraviolet sensitivity of the soft terms altogether. If
such a program succeeds, there are two immediate advantages. First of all,
one has control over the flavour violations in the soft terms. Moreover,
in this case, the soft terms are necessarily computable ({\it i.e.} their
quantum corrections are finite in the effective theory below $M_{Pl}$) and
therefore one can make definite mass predictions relevant to experimental
searches.

The best known class of theories in which the soft terms are insensitive 
to the far ultraviolet is given by gauge-mediated models~\cite{gaug}. 
Here the original supersymmetry breaking is felt at tree level only by some
new particles of mass $M$ (the messengers) and then communicated to the
observable sector by loop diagrams involving ordinary Standard Model gauge
interactions. Quantum corrections to the soft terms vanish for momenta
larger than $M$, as schematically illustrated in fig.~1 by the lines indicated
with the caption ``Gauge mediation". 
Any dynamics occurring at energy scales above
$M$ do not affect the soft terms. If we assume that $M$ lies below any
new flavour dynamics, then the Yukawa couplings provide the only source
of flavour violations and we recover a  supersymmetric extension of the
GIM mechanism. Flavour violations in low-energy hadronic and leptonic
processes are fully under control.

In gauge mediation, the soft terms are finite and computable. In the simplest
version of the model, the gaugino, squark, and slepton masses are given by
\beqa
{\tilde m}_{g_i} &=& \frac{\alpha_i F}{4\pi M},\\
{\tilde m}_f^2 &=& 2\sum_{i=1}^3 C_f^i \left( \frac{\alpha_i F}{4\pi M}
\right)^2 .
\eeqan
Here $\alpha_i$ are the Standard Model gauge coupling constants and 
$C_f^i$ are the corresponding quadratic Casimir coefficients.

Recently a different approach to obtain ultraviolet insensitivity of the
soft terms has been pursued. The central observation is that, in the presence
of supersymmetry breaking, gravity generates soft terms even if there are
no direct
couplings between the hidden and observable sectors~\cite{lis,noi}. 
This is an effect
of the superconformal anomaly and it gives rise to soft terms that are 
suppressed by loop factors. If tree-level soft terms exist, then the 
anomaly-induced terms are subdominant. However, in some cases, they can
provide the leading contribution. For gauginos, this occurs when the theory
does not contain any gauge-singlet superfield that breaks supersymmetry
(as for theories with dynamical supersymmetry breaking)~\cite{noi}. Indeed,
in the absence of gauge singlets $X$, one cannot generate the gaugino masses
${\tilde m}_g$ from the usual operator
\beq
\int d^2\theta \frac{X}{M_{Pl}}~{\rm Tr} W^\alpha W_\alpha +{\rm h.c.},
\eeqn
and therefore one has to rely on higher-dimensional operators,
which give at most ${\tilde m}_g \sim F^{3/2}/M_{Pl}^2 \sim$~keV.
For scalars, the absence of tree-level contributions to their soft masses
can be obtained with specific structures of K\"ahler potentials. These
structures occur when the supersymmetry-breaking and observable sectors
reside on different branes embedded into a higher-dimensional space and
separated by a sufficiently large distance~\cite{lis}.

Let us assume that the soft terms, for the reasons 
explained above (or for any other
unknown reason), are dominated by the anomaly contribution.
In this case, the gaugino masses are given by~\cite{lis,noi}
\beq
{\tilde m}_g =\frac{\beta_g }{g}~m_{3/2},
\eeq{gau}
where $m_{3/2}$ is the gravitino mass (a measure of the supersymmetry-breaking
scale) and $\beta $ is the corresponding gauge-coupling beta function. More
explicitly, for the gauginos relative to the three factors of the
Standard Model gauge group, eq.~\leqn{gau} gives
\beqa
M_3 &=& -\frac{3\alpha_s}{4\pi}~m_{3/2} \CR
M_2 &=& \frac{\alpha}{4\pi \sin^2\theta_W}~m_{3/2}\simeq -0.1 M_3\CR
M_1 &=& \frac{11\alpha}{4\pi \cos^2\theta_W}~m_{3/2}\simeq -0.3 M_3.
\eeqa{anom}
This is to be compared with the usual gaugino mass relations under
GUT assumptions, ${\tilde m}_g =(g^2/g_{GUT}^2){\tilde m}_g(M_{GUT})$,
which give 
\beqa
M_2 &= & 0.30~ M_3 \CR
M_1 &= & 0.17~ M_3. 
\eeqa{ggut}

The anomaly-mediated mass relation in eq.~\leqn{gau} is particularly
interesting because it depends only on low-energy coupling constants and
it makes no reference on high-energy boundary
conditions (GUT, messengers, ...). 
Indeed the form of eq.~\leqn{gau} is invariant under 
renormalization group transformations. This entails a large degree of
predictivity, since all soft terms can be computed from known low-energy
Standard Model parameters and a single mass scale, $m_{3/2}$. Also, it
leads to robust predictions, since the renormalization group invariance
guarantees complete insensitivity of the soft terms from ultraviolet
physics. As demonstrated with specific examples in ref.~\cite{noi},
heavy states do not affect the soft terms, since their contributions to the
$\beta$ functions and to threshold corrections exactly compensate each other.
This means that the gaugino mass predictions in eqs.~\leqn{anom} are
valid irrespective of the GUT gauge group in which the Standard Model
may or may not be embedded\footnote{However, exceptions to ultraviolet 
insensitivity appear in the presence of gauge-singlet superfields~\cite{ricc}.}.
Therefore, although the soft terms are generated at very high-energy scales,
their renormalization group trajectories are determined in such a way that
the low-energy values of the soft terms are specified only by low-energy
parameters. This is schematically illustrated in fig.~1 by the lines
indicated with the caption ``Anomaly mediation". Whatever the
dynamics that
breaks flavour symmetry at high energies may be, the low-energy
soft terms will respect
a super-GIM mechanism.

In spite of its great theoretical appeal, a supersymmetric model with
anomaly-mediated mass spectrum is not phenomenologically acceptable. The
problem lies in the form of the scalar masses~\cite{lis}
\beq
{\tilde m}^2 =-\frac{1}{4} \left( \frac{\partial \gamma}{\partial g}
\beta_g + \frac{\partial \gamma}{\partial y}\beta_y \right) m_{3/2}^2.
\eeq{scala}
Here $\beta_g$ and $\beta_y$ are the beta functions for the gauge and
Yukawa coupling $y$, and $\gamma$ is the anomalous dimension of the
corresponding superfield. In the supersymmetric model $SU(3)$ is asymptotically
free and has a negative $\beta$ function, but $SU(2)$ and $U(1)$ have a
positive $\beta$ function. Therefore, eq.~\leqn{scala} predicts positive
squark squared masses, but negative slepton squared masses. This would induce
a spontaneous breaking of QED.

Several possible solutions have been suggested in order to cure this 
problem~\cite{lis,ricc,alt}. All of these solutions of course require new positive 
contributions to the slepton masses. These new terms necessarily spoil the most 
attractive feature of anomaly mediation, {\it i.e.} the renormalization group
invariance of the soft terms and the consequent ultraviolet insensitivity.
This is the most disappointing aspect of this scenario. At present, it is
too early to assess if some of the appealing features of anomaly mediation
have any relevance in the description of the elementary particle world.

\subsection{Experimental Consequences}
  
The realization that there are
several possible schemes of supersymmetry-breaking communication
has profound experimental implications, not only because of
the different patterns of the superpartner mass spectrum, but also because
each scheme has very distinctive signatures at high-energy collisions.
Therefore, the search for supersymmetry requires different experimental 
analyses aimed at identifying quite different signals. 

The stereotype
missing-energy signature of supersymmetry is specific to gravity-mediated
scenarios, in which the produced supersymmetric particles cascade decay
into the invisible lightest neutralino. 

In gauge-mediated scenarios,
the gravitino is the lightest supersymmetric particle, because its mass
is determined by gravitational interactions instead of gauge interactions
as in the case of the other superpartners. The experimental signals
are then determined by the nature of the next-to-lightest supersymmetric
particle (either a neutralino or a stau, depending on model-dependent 
parameters) and the scale of supersymmetry breaking $F$ (which determines
the lifetime of the next-to-lightest supersymmetric
particle). 
For $\sqrt{F}\lsim 10^6$~GeV, the next-to-lightest supersymmetric
particles promptly decay into their Standard Model partners and
gravitinos, leaving topologies containing tau leptons and missing
energy (in the case of the stau) or photons and missing
energy (in the case of the neutralino). On the other hand, 
for $\sqrt{F}\gsim 10^6$~GeV, the next-to-lightest supersymmetric
particle is quasi-stable, since its decay length is typically longer than
the detector size. The experimental signature is given by missing energy
in the case of the neutralino, while in the case of the stau there is a
more unconvential signal coming from a heavy charged particle crossing the
apparatus, leaving anomalous ionization tracks.

The gaugino mass relations in eqs.~\leqn{anom}, characteristic of anomaly
mediation, lead to quite peculiar experimental signals. Indeed, 
eqs.~\leqn{anom} predict $M_2<M_1$ (in contrast to the usual case
of eqs.~\leqn{ggut}, in which $M_1<M_2$). This and the electroweak-breaking
conditions imply that, in realistic
models, the $SU(2)$ $W$-ino triplet is almost degenerate in mass. 
The mass splitting inside the triplet
is dominated by loop effects and the charged particle is heavier
than the neutral one, with $m_{\chi^\pm}-m_{\chi^0}$ 
in the range between
the pion mass and about 1 GeV~\cite{lis2,noi2}.  
The (mainly $W$-ino) neutralino is the
lightest supersymmetric particle, and the first chargino decays into
a neutralino and a relatively
soft pion $\chi^\pm \to m_{\chi^0} \pi^\pm$. The experimental difficulty 
lies in triggering such events, although kinks in the vertex detector could
be revealed at the analysis stage. Different strategies consist in tagging
high-energy jets or photons~\cite{lis2} or focus on the production and decay
of other supersymmetric particles~\cite{noi2}.

From this brief discussion, it should be clear that very different experimental
strategies and analyses are necessary to look for the diversified ways in 
which supersymmetry could reveal itself in high-energy collisions.

As we have previously discussed, the 
various schemes of supersymmetry-breaking communication differ in the
way they address the flavour problem. Therefore it is not surprising that
experiments searching for rare flavour-violating or CP-violating processes
are of great value for discriminating between the different supersymmetric 
scenarios.
We can distinguish between two classes of supersymmetry-breaking models:
{\it i)}
those (like gauge mediation or anomaly mediation with a universal extra
contribution to scalar masses) which satisfy a super-GIM mechanism, and 
flavour or CP violation is originating only from CKM angles and phases;
{\it ii)} those (like gravity mediation) which rely on some flavour symmetry
valid at some very large scale in the proximity of $M_{Pl}$, and necessarily
contain some new sources of flavour and CP violation in the 
supersymmetry-breaking parameters. 

In models belonging to class {\it i)}, we can expect only rather moderate
deviations from the Standard Model predictions in flavour processes.
The only exceptions could come from processes that are accidentally
suppressed in the Standard Model and are 
more sensitive to new physics corrections (as in the case of
the rare decay $B\to X_s
\gamma$).
On the other hand, in the models of class {\it ii)}, it appears almost
unavoidable that new flavour-violating and CP-violating effects should
lurk just behind the present experimental limits~\cite{hall}. In this respect,
the r\^ole of B factories will be crucial in helping theorists to sort out
the way in which supersymmetry breaking is realized. Similarly, improvements
in the sensitivity on lepton-family 
violating processes (like $\mu \to e \gamma$ and 
$\mu$--$e$ conversion
in nuclei) and CP-violation (like electron and neutron electric dipole
moments) will bring very valuable information.

Recently, the KTeV~\cite{ktev} and NA48~\cite{na48} collaborations have 
announced new 
results  for $\epsilon{'} /\epsilon$,
leading to a world
average~\cite{dag} of 
Re~$\epsilon{'} /\epsilon =(21.4\pm 4.0)\times 10^{-4}$. This
value is higher than the predictions made within the Standard Model~\cite{eps},
and stirred some interest 
on the possibility that supersymmetric
effects had been observed~\cite{gabr,mura,epss,math}. 

However, it is not impossible for the Standard Model to accommodate the
measured value of $\epsilon{'} /\epsilon$. 
For instance, this can be done by taking the hadronic parameter $B_6$ 
to be about 1.5. This moderate enhancement of $B_6$ 
with respect to the traditional
expectations  is not unreasonable. Large contributions to $B_6$ 
are found from ${\cal O}(p^2/N_c)$ corrections in the $1/N_c$ 
expansion~\cite{dort}
and in the chiral quark model~\cite{trie}.
This could be the result of a $\Delta I =1/2$ rule for the operator $Q_6$,
analogous to the one that applies to the operators $Q_1$ and $Q_2$.
It has also been found that isospin-violating effects arising from
the mass difference between up and down quark~\cite{germ} and final-state
interactions~\cite{pich} both contribute to increasing the
estimate of $\epsilon{'} /\epsilon$. 

Therefore, it appears likely that the discrepancy between theory and
experiment is just caused by our present poor knowledge of the hadronic matrix
elements. Nevertheless, it is interesting to wonder whether supersymmetry
can be responsible for a significant enhancement of the prediction for
$\epsilon{'} /\epsilon$.

In models of class {\it i)}, where the flavour and CP violations 
originate purely  
from CKM effects, supersymmetric contributions to $\epsilon{'} 
/\epsilon$ are very small and, moreover,
in general they
tend to reduce the Standard Model prediction~\cite{gabr}. In models
of class {\it ii)}, the new sources of flavour violations are usually
parametrized by (complex) flavour non-diagonal entries in the squark
mass matrices. Constraints from $\Delta m_K$ and $\epsilon$ imply that
flavour-violating mass insertion in the left or in the right squark sectors
cannot give significant enhancements to $\epsilon{'} /\epsilon$. On the 
other hand, a mass insertion mixing left and right squarks is less
constrained and it can give a contribution to
$\epsilon{'} /\epsilon$ of the size of the measured value.
It is interesting that one does not need to rely on unexpectedly large
left--right squark mixings to obtain this result. Indeed, the experimental
result can be explained with a
``theoretically reasonable"
guess for the flavour-violating left--right mixing~\cite{mura}
\beq
{\tilde m}_{d_L s_R}^2 \sim m_{3/2} m_s \sin \theta_c e^{i\delta}.
\eeqn
Here $m_{3/2}$ is the typical supersymmetry-breaking mass, $m_s$ is the
strange quark mass, $\theta_c$ is the Cabibbo angle, and $\delta$ is a
phase of order 1. The use of similar ``reasonable" estimates for the squark
and slepton mass matrices leads to distinctive predictions, which will allow
us to
test these assumptions. Indeed, the neutron electric
dipole moment and the branching ratio for $\mu \to e \gamma$ should lie
just beyond the present experimental limits.

Recently, Kagan and Neubert~\cite{math} 
have made the very interesting observation
that, in the presence of mass splittings between the squarks ${\tilde u}_R$
and ${\tilde d}_R$, gluino box diagrams can generate $\Delta I =3/2$
amplitudes that are enhanced by the $\Delta I =1/2$ selection rule. This
gives a potentially very large effect on $\epsilon{'} /\epsilon$, which
can explain the experimental result even for squark masses in the TeV region.

\section{Extra Dimensions}

One of the greatest scientific successes of the last twenty years has been
the precise verification of the Standard Model as the correct theory 
describing elementary particle interactions up to the weak scale. Following
the idea of grand unified theories, we are used to extrapolating our
knowledge to much smaller length scale, of the order of
$M_{GUT}^{-1}\sim 10^{-32}$~m. Moreover, string theory suggests a way to unify
gauge and gravity forces at an even smaller distance scale,
$M_S^{-1}\sim 2/(\sqrt{k\alpha_{GUT}}M_{Pl})$.
Figure 2 illustrates
the presumed behaviour of the gauge and gravitational couplings emerging
from these conjectures. 

\begin{figure}
\begin{center}
\epsfig{file=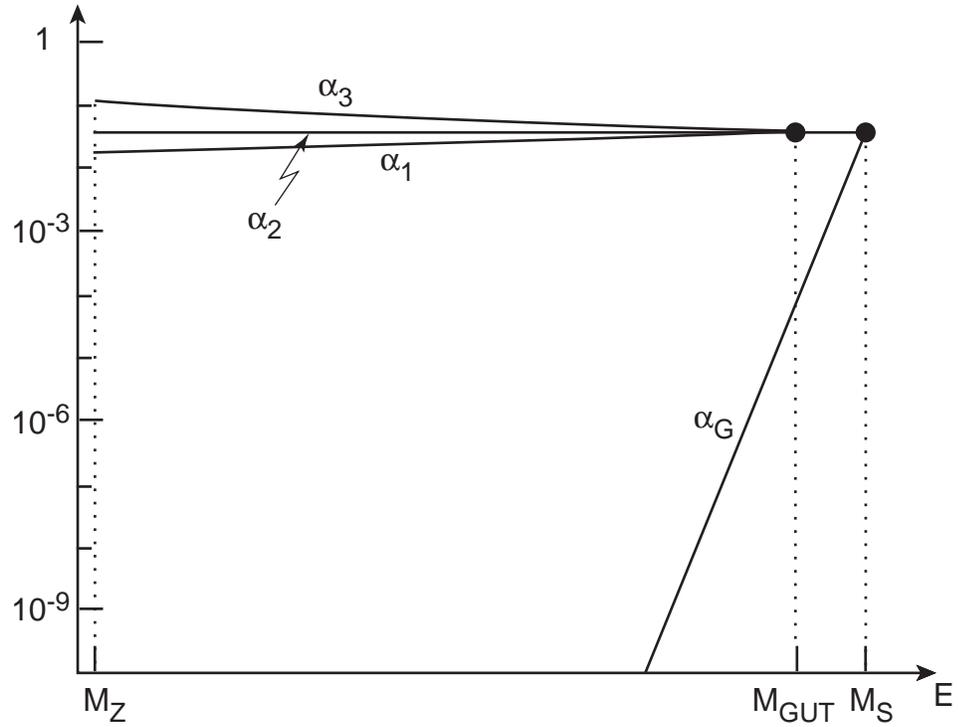}
\caption{The behaviour of the three gauge coupling constants
and the gravitational coupling $\alpha_G \sim E^2/M_{Pl}^2$, as a function
of the energy $E$ in the traditional scenario with grand unification at
the scale $M_{GUT}$ and superstrings at the scale $M_S$.}
\end{center}
\end{figure}

These are certainly courageous theoretical extrapolations, but nevertheless 
are not
at present experimentally confirmed. In particular, gravity has been
experimentally tested only up to scales of the order of $\lambda \sim {\rm mm}
\sim (2\times 10^{-4}~{\rm eV})^{-1}$, {\it i.e.} 30 orders of magnitude larger
than $M_S^{-1}$! 
It is therefore legitimate to question the scenario illustrated
in fig.~2, and wonder whether the gravitational coupling $\alpha_G$ could
evolve, at energies above $\lambda^{-1}$, 
quite differently from our traditional expectations.
In particular, one could imagine that the gravitational coupling becomes
of the order of the gauge couplings already at the weak scale, eliminating
the need for the large mass parameter $M_{Pl}$ or, in other words, eliminating
the notorious hierarchy problem.

Arkani-Hamed, Dimopoulos, and Dvali~\cite{add} 
have suggested a physical setting in
which this radical point of view can actually be realized. Their
construction assumes that our 4-dimensional world, in which ordinary
particle processes occur, is actually embedded into a higher-dimensional
space, in which only gravitons are free to roam. 
Let us define the total number of dimensions as $D=4+\delta$ and assume that
the $\delta$ extra dimensions are compactified in a space with volume 
$V_\delta$. 
It is a simple geometrical
exercise to prove that the effective Newton constant in the 4-dimensional
theory is related to the fundamental energy scale $M_D$ of the full
$D$-dimensional gravitational theory by the equation
\beq
G_N^{-1}\equiv M_{Pl}^2 =M_D^{2+\delta}V_\delta .
\eeq{newt}
From this, we infer the typical radius of the compactified space
\beq
R\sim V_\delta^{1/\delta} \sim \frac{1}{M_D} \left( \frac{M_{Pl}}{M_D}
\right)^{2/\delta} .
\eeq{rad}
If we want to realize the scenario in which the fundamental gravitational
mass parameter is roughly equal to the weak mass scale, we have to 
insist that $M_D \sim$~TeV, and therefore the typical size of the
compactification radius is
\beqa 
R= (5\times 10^{-4}~{\rm eV})^{-1}\sim 0.4~{\rm mm}~~~~~~~&{\rm for}~
\delta=2,& \CR
R= (20~{\rm keV})^{-1}\sim 10^{-5}~\mu{\rm m}~~~~~~~&{\rm for}~
\delta=4,& \CR
R= (7~{\rm MeV})^{-1}\sim 30~{\rm fm}~~~~~~~&{\rm for}~
\delta=6.&
\eeqa{radd}
For $\delta =1$ the size of $R$ is of astronomical length and therefore
excluded by standard 
observations. The case $\delta =2$ is marginally allowed and
therefore interesting for experiments aiming at improving gravitational
tests at small distances. As $\delta$ grows, $R$ approaches the inverse
of the fundamental mass scale $M_D$. 

Before proceeding, we have to discuss whether the construction of 
ref.~\cite{add}
can be realized in a physical system. Localizing fields on subspaces with
lower dimensions can be achieved in a field theoretical context,
but requires the introduction of certain scalar fields with particular
potential; it is therefore possible but not straightforward. 
The great interest in the proposal of ref.~\cite{add} has been
stirred by the observation that this situation is rather generic in the
context of string theory. Indeed, Dirichlet branes (the space defined
by the end-points of open strings~\cite{brane}) 
are defects intrinsic to string theory
on which the gauge theory is confined. The picture of ordinary particles
(open strings) localized on the brane with gravity (closed strings)
propagating in the bulk can be realized in string models~\cite{mode}.
This observation could actually help in bringing closer two lines of
research in theoretical physics (one more phenomenologically oriented and
one more formally oriented), which seemed to follow different paths in the
last years. Indeed many theoretical speculations intended for Planck
energy scales could now be relevant at the TeV scale, and therefore 
experimentally tested.

As evident from 
eq.~\leqn{rad}, 
in the higher-dimensional context, the weakness of gravity or, in other
words, the smallness of the ratio $G_N/G_F$ is related to the largeness
of the number $RM_D$, which measures the compactified radius in its natural
units.
The hierarchy problem is not completely solved unless we understand
why $R^{-1}\ll M_D \sim$~TeV. 
There have been several attempts to find dynamical explanations for the
radius stabilization~\cite{stab}. This problem may be connected with the 
cosmological constant puzzle.

Around the time of this Conference, many new ideas in theories with
extra dimensions are being proposed. Some of them are very interesting
alternatives to the scenario of ref.~\cite{add} as a solution to the hierarchy
problem.

Randall and Sundrum~\cite{rs1} have proposed a higher-dimensional scenario
in which the hierarchy problem is solved without the need for large
($R\gg M_D^{-1}$) extra dimensions. They consider a 5-dimensional
non-factorizable geometry ({\it i.e.} 
the 4-dimensional metric is not independent
of the extra coordinates) in which the line element is given by
\beq
ds^2=e^{-2kr_c\Phi} \eta_{\mu \nu} dx^\mu dx^\nu +r_c^2 d\Phi^2.
\eeq{metr}
Here $k$ is an energy scale of the order of the 5-dimensional Planck mass
$M_5$ and $\Phi$ ($0\le \Phi \le \pi$) is the coordinate of the compactified
extra dimensions of size $r_c$. This metric is the solution of the Einstein
equation in a model with two 3-branes (at $\Phi=0$ and $\Phi=\pi$)
with opposite tensions tuned to
preserve 4-dimensional Poincar\'e invariance. In this situation, the 
4-dimensional Planck mass is given by
\beq
M_{Pl}^2=\frac{M_5^3}{k} \left( 1-e^{-2\pi kr_c}\right).
\eeq{plc}
We will be interested in the limit $kr_c\gg 1$, in which the exponential
factor in eq.~\leqn{plc} is irrelevant, and we take $M_5\sim k\sim {\cal O}
(M_{Pl})$. The exponential factor is however important for the mass
parameters of the fields confined on the 3-brane at $\Phi =\pi$
representing our world.
As apparent from eq.~\leqn{metr}, the exponential $e^{-2kr_c\Phi}$ acts as
a conformal factor in the 4-dimensional theory and therefore it is not
surprising that the physical mass parameters on the brane are given by
$m_0e^{-\pi k r_c}$, if $m_0 \sim {\cal O}(M_{Pl})$ is the mass parameter
in the 5-dimensional theory. For the moderate number $kr_c \simeq 50$,
the large hierarchy between the weak and the gravitational
mass can be reproduced.

The emerging physical picture is the following. Because of the non-factorizable
form of the geometry, the gravitational field configuration is highly
non-trivial. Gravitons are localized on one brane, while the Standard
Model particles live on the other brane. The small overlap of the graviton
wave-function with our brane explains the weakness of gravity. No hierarchically
small numbers are required because of the exponential suppression. The
mass gaps and the mass scale in the effective interactions of the Kaluza--Klein
gravitons are both of the order of the weak scale, since the weak scale is
the only relevant mass in this physical picture.

This proposal has been further elaborated and an alternative scenario for
a solution to the hierarchy problem has been suggested in ref.~\cite{rl}.
The crucial observation~\cite{rs2} is that, in the presence of non-factorizable
metrics we can envision non-compact extra dimensions without conflicting
with observations. The graviton is again localized, but its Kaluza--Klein 
spectrum has no mass gap. Nevertheless this is not problematic, because
all excited Kaluza--Klein modes give corrections to the gravitational
couplings of the order of $E^2/M_{Pl}^2$, where $E$ is the typical process
energy. It is now possible to consider a setup in which the Standard
Model resides on one brane while gravity is localized on a different
brane, and both branes have positive tensions. The separation between
the two branes reproduces the hierarchy $M_W/M_{Pl}$ and the fifth
dimension is infinitely large.

Another very interesting proposal was recently suggested by Cohen and 
Kaplan~\cite{cohe}. 
They consider a 6-dimensional setup consisting of  gravity and 
one scalar field $\Phi$, with a scalar potential that allows a 3-dimensional
global ``cosmic string" solution. The string core is identified with our
4-dimensional space-time. After solving the Einstein equations for this
system, one finds that the effective Planck mass in 4 dimensions is given by
\beq
M_{Pl}^2=\pi \Gamma (3/8)~\left( \frac{M_6}{f} \right)^{9/2}~
e^{(M_6/f)^4} ~M_6^2.
\eeq{ck}
Here $M_6$ is the fundamental mass of the underlying 6-dimensional theory
and $f$ is the asymptotic vacuum expectation value of the scalar field
$\Phi$. A ratio $M_6/f \sim 2.7$ is sufficient to generate the large hierarchy
between the weak and gravitational scale, because of the steep functional
dependence ($\sim e^{x^4}$) in eq.~\leqn{ck}. The resulting effective theory
looks very similar to the one proposed in ref.~\cite{add}, but the hierarchy
$M_W/M_{Pl}$ is now dynamically explained.

\subsection{Opening New Problems}

The idea of having a unique fundamental mass scale, of the order of the TeV,
for both weak and gravity interactions clearly requires a complete rethinking
of much of the accepted understanding of the high-energy behaviour and of 
early cosmology. 

First of all, one has to abandon a very successful feature of the traditional
constructs: certain symmetry-breaking interactions are small because they
arise from physics at very large scales. Usually one describes these 
symmetry-breaking effects with effective operators suppressed
by an unspecified mass scale $\Lambda$, such as
\beqa
{\rm neutrino~masses}&\to& \frac{1}{\Lambda} \ell \ell HH \CR
{\rm proton~decay}&\to& \frac{1}{\Lambda^2} qqq\ell \CR
{\rm flavour~violation}&\to& \frac{1}{\Lambda^2} \bar s d \bar s d \CR
{\rm lepton~family~violation}&\to& \frac{1}{\Lambda} \bar \mu \sigma_{\mu 
\nu} e F^{\mu \nu}. 
\eeqa{list}  
The smallness of the observed violation of the corresponding
exact or approximate
symmetries implies that the scale $\Lambda$ is much larger than the weak
scale. In theories with quantum gravity at the TeV scale, we cannot rely on such
an explanation. These theories therefore require new mechanisms to
understand small parameters. One possibility is that small parameters are
not the consequence of approximate symmetries, as in the examples above,
but instead in what I will call ``locality and geometry". As suggested
in ref.~\cite{flav}, suppose that all unwanted symmetry-breaking effects can
only occur locally on branes that are physically separated by a distance
$d$ from the 3-dimensional brane of our world. In this case, the effective
couplings of the symmetry-breaking interactions will be suppressed by
a factor $e^{-m/d}$, where $m$ is the typical mass of the bulk particle
that mediates the interaction from one brane to the other. Large suppression
factors can be obtained with moderate ratios of $m/d$. 

The same mechanism can be used to obtain the flavour structure of the
Yukawa couplings~\cite{yukk}. 
One can also extend
this picture and place the three quark and lepton families on different
locations in the directions orthogonal to the ordinary 3-dimensional 
space~\cite{yukkk}. 
Depending on the profile of the fermion wave-functions along the
extra dimensions, large hierarchies in the Yukawa couplings could
be obtained from numbers of order 1, using the above-mentioned
exponential factor. 
If this conjecture were true, we could even hope to unravel unsuspected
properties of the flavour symmetries. The pattern of Yukawa couplings
could look much simpler when viewed in terms of exponential factors or
some other functional dependence. 

Neutrino masses cannot be any longer explained by the see-saw mechanism and
require some new higher-dimensional mechanism. One possibility is that
right-handed neutrinos, in contrast with the other Standard Model particles,
live in the full $D$-dimensional space~\cite{neutrhd}. The Yukawa interaction 
between left- and right-handed neutrinos is localized on the brane. Since
the wave-function of $\nu_R$ is spread in the bulk space, the effective
Yukawa coupling is suppressed by the square root of the compactified
volume $V_\delta$. The neutrino mass is then given by
\beq
m_\nu=\frac{\lambda \langle H \rangle}{\sqrt{V_\delta M_D^\delta}}
\sim \lambda \langle H \rangle \frac{M_D}{M_{Pl}} \sim 10^{-4}~{\rm eV}
\left( \frac{M_D}{\rm TeV}\right) ,
\eeq{nms}
where we have assumed that the Yukawa coupling $\lambda$ in the $D$-dimensional
theory is of order 1. Notice that the resulting neutrino mass is of the
Dirac type and it is in the correct ballpark to explain the atmospheric
neutrino data.

Although it first appears that gauge-coupling unification
is irremediably lost, it is nevertheless possible to
conceive new higher-dimensional schemes in which the success of the
supersymmetric prediction is recovered.
One possibility~\cite{gauc} is to assume that Standard Model particles have
Kaluza--Klein excitations (with masses larger than a few TeV). Their effects 
in the $\beta$ functions change the logarithmic dependence 
on the energy into a power dependence and speed up the unification,
which can now occur at energies not much larger than the weak scale. From the
field-theoretical point of view, one loses control of the theory, but 
nevertheless it is possible that an actual gauge-coupling unification 
is achieved in a string theory with TeV scale. Another possibility~\cite{gauc2}
is to use field variations in the large extra dimensions to
achieve a logarithmic unification. 

The early cosmology of theories with quantum gravity at the TeV scale
will also look drastically different from what has been traditionally
assumed. In the scenario of ref.~\cite{add}, a problem arises. During the
early phase of the Universe, energy can be emitted from the brane into the bulk
in the form of gravitons. The gravitons propagate in the extra dimensions
and can decay into ordinary particles only by interacting with the brane, and
therefore with a rate suppressed by $1/M_{Pl}^2$. Their contribution to
the present energy density exceeds the critical value unless~\cite{add}
\beq
T_\star < \frac{M_D}{\rm TeV}~ 10^{\frac{6\delta -15}{\delta +2}}{\rm MeV}.
\eeq{tstar}
Here $T_\star$ is the maximum temperature 
to which we can simply extrapolate the thermal history of
the Universe, assuming it is in a 
stage with completely stabilized $R$ and with vanishing energy density
in the compactified space. 
As a possible example of its origin, 
$T_\star$ could correspond to the reheating temperature
after an inflationary epoch.
The bound in
eq.~(\ref{tstar}) is very constraining. In particular,
for $\delta =2$, only values of $M_D$ larger than about 6 TeV can lead
to $T_\star > 1$ MeV and allow for standard nucleosynthesis. Moreover,
even for larger values of $\delta$, eq.~(\ref{tstar}) is
very problematic for any
mechanism of baryogenesis~\cite{dav}. 

The graviton emission is also dangerous in an astrophysical context.
Extra-dimensional gravitons would speed up supernova cooling in contradiction
with the neutrino observation from SN1987A, unless~\cite{culle}
\beqa
M_D >50~{\rm TeV}~~~~ &{\rm for~}& \delta=2, \CR
M_D >4~{\rm TeV}~~~~ &{\rm for~}& \delta=3.
\eeqa{cul}
An even stronger limit comes from distortion of the diffuse gamma-ray 
background~\cite{halgr},
\beqa
M_D >110~{\rm TeV}~~~~ &{\rm for~}& \delta=2, \CR
M_D >5~{\rm TeV}~~~~ &{\rm for~}& \delta=3.
\eeqa{hal}
This bound is very constraining in the case of two extra dimensions, and it 
rapidly decreases with $\delta$, because of the power-law suppression
of graviton interactions. Notice that these limits are determined by the
infrared behaviour of the gravitational theory. Therefore they do not
apply to theories that have large Kaluza--Klein graviton gaps. 
They can also be evaded in the scenario of ref.~\cite{add}, in the case 
of very particular compactified spaces which enhance the masses of the first
Kaluza--Klein excitations.

\subsection{Experimental Tests}

The idea that quantum gravity resides at the weak scale can be put under
experimental scrutiny. We started our discussion on the motivations of
extra dimensions by pointing out that gravity has been tested only to
scales just below the millimetre. It is therefore clear that improvements
in the experimental sensitivity will be of great importance. Indeed there
are ongoing experiments~\cite{pirl} that aim at testing gravity up to distances
of several tens of microns.
 
Unfortunately, the astrophysical bounds presented in  eq.~\leqn{hal}
can be translated into a limit on the Compton wavelength of the first
graviton Kaluza--Klein mode of $5\times 10^{-2}~\mu$m. The possibility of 
experimentally observing a deviation of gravity caused by higher-dimensional
gravitons is then
ruled out, at least in near-future experiments. Any modification
of the compactified space capable of avoiding the astrophysical bound will
also exclude a visible signal at short-distance gravitational experiments.
Nevertheless, in many models realizing the idea of low-scale quantum gravity,
there exist other light bulk particles, which could lead to observable
signals~\cite{add}. 
A possible effect could also come from other light particles in
scenarios with low-energy supersymmetry breaking~\cite{sav}.

High-energy collider experiments can directly probe the new dynamics
of quantum gravity at the weak scale. At first, one may believe that
the experimental signal should depend on the specific quantum gravitational
theory, and therefore no solid prediction could be made. However, in 
a certain kinematical regime, it is possible to make rather model-independent
estimates of graviton production in high-energy collisions. 
The strategy is to use an effective theory~\cite{giu,pesk}, valid below the
fundamental mass scale $M_D$, where one can perform an expansion in $E/M_D$
(here $E$ is the typical process energy) and use our knowledge of the
infrared properties of gravity.

In the scenario of ref.~\cite{add}, gravitons are massless particles
propagating in $D$ dimensions. Therefore, the relation between their energy $E$
and their momentum is $E^2={\vec p}^2+p_{extra}^2$, where $\vec p$ describes
the usual 3-dimensional components and $p_{extra}$ is the momentum along
the extra dimensions. This relation gives an intuitive explanation of
how a $D$-dimensional particle can
be described by a collection of 4-dimensional modes (called the Kaluza--Klein
excitations) with mass $m=|p_{extra}|$. 

We will be interested in the 
production of the Kaluza--Klein graviton modes in high-energy collisions.
The single production of a graviton with non-vanishing $|p_{extra}|$ violates
momentum conservation along the extra dimensions. This is not surprising,
since the presence of the 3-brane breaks translational invariance in the
directions orthogonal to the brane.  It is like playing tennis against a
wall: the momentum along the direction
orthogonal to the wall is not conserved. Gravitons cannot be directly
detected.
Therefore the signal in collider experiments is missing energy and imbalance
in final-state momenta, caused by the graviton escaping in the extra-dimension
compactified
space. Just for illustration, we can 
visualize elementary-particle interactions as the collisions
of balls 
on a pool table. The balls can only move on a  2-dimensional surface (the
brane), but as they knock each other they can emit a sound wave (the graviton),
which travels
in the air (the bulk). Because of this energy loss, an observer living on 
the surface of the table can infer the existence of the extra dimension
by measuring the kinematics of the balls before and after the collision.

Each graviton Kaluza--Klein mode $G_n$ has a production probability proportional
to $E^2/M_{Pl}^2$, which gives rise to a cross section at hadron colliders of
\beq
\sigma (pp\to G_n ~
{\rm jet})\simeq \frac{\alpha_s}{\pi} ~G_N =10^{-28}~{\rm fb}.
\eeqn
This is hopelessly small and it cannot be observed. However, experiments are
sensitive to inclusive processes, in which we sum over all kinematically
accessible Kaluza--Klein modes. Because of the large volume in the extra
dimensions, the number of graviton  Kaluza--Klein modes with mass less than
a typical energy $E$ is very large $\sim E^\delta M_{Pl}^2/M_D^{2+\delta}$.
As a result, the dependence of the inclusive cross section on $M_{Pl}$
cancels out, and we find
\beq 
\sum_n \sigma (pp\to G_n {\rm jet})\simeq \frac{\alpha_s}{\pi} ~
\frac{E^\delta}{M_D^{2+\delta}}.
\eeqn

By studying final states with photons and missing energy, LEP has already
set bounds on the fundamental quantum gravity scale $M_D$ of about
1 TeV (for a number of extra dimensions $\delta =2$)~\cite{franc}. 
Future studies at the Tevatron, LHC, 
linear colliders or muon colliders can significantly extend the sensitivity
region of $M_D$ by analysing final states with jets and missing 
energy or photons and missing energy~\cite{giu,pesk}.

It should be stressed that in a complete quantum gravity theory there will
certainly exist other experimental signals, quite different from
the graviton signal considered above. However, these new signatures are
model-dependent and cannot be predicted without a complete knowledge
of the final theory. Therefore, the effective-theory signal discussed
here, although it does not necessarily represent the discovery mode, is
best suited for setting reliable bounds on $M_D$.

In general, one can parametrize new physics effects at the scale $M_D$ with all
possible effective interactions with couplings of order 1 in 
units of $M_D$. However, there is one particular operator that could
play a special role,
\beq
{\cal T}\equiv T_{\mu \nu}T^{\mu \nu}-\frac{1}{\delta +2} T^\mu_\mu T^\nu_\nu.
\eeq{opt}
Here $T_{\mu \nu}$ is the energy--momentum tensor. The operator in 
eq.~\leqn{opt} is induced by tree-level virtual graviton exchange 
and it will appear in the effective Lagrangian with
a coefficient of order $1/M_D^4$. Unfortunately
the precise form of this coefficient cannot be computed by using only the
effective theory, because it depends 
on ultraviolet properties.
Nevertheless, experimental searches on the existence of this operator are
interesting because they represent a test on the spin-2 nature of the
particle that mediates the effective interactions. The operator ${\cal T}$
gives rise to a variety of experimental signals, which include, in $e^+e^-$
colliders, $d$-wave contributions to fermion pair production, $\gamma
\gamma$ and multijet final states and, in hadron colliders, dilepton or
$\gamma \gamma$ production~\cite{giu,hew}. All these signals are in principle
related, because they originate from the same interaction.

The graviton-production signal is characteristic of theories with large
extra dimensions $R\gg M_D^{-1}$. In models in which the graviton 
Kaluza--Klein gaps are of the order of $M_D$ (as for instance in the
scenario of ref.~\cite{rs1}),
the interesting experimental signal is given by the production  
of the new gravitational
excitations with weak-scale masses. Actually, it is possible
that all Standard Model particles have Kaluza--Klein modes at the TeV
scale~\cite{anto}. This is the case, for instance, in the proposal of
ref.~\cite{gauc} to achieve gauge-coupling unification at low-energy scales.
This situation is not inconsistent with the large extra dimension scenario.
The Standard Model could live in a $D'$-dimensional space with
$4<D'<D$ and with compactification radius $R'\sim$~TeV. Gravity propagates
also in the extra $D-D'$ dimensions characterized by a radius $R\gg R'$.
Precision electroweak measurements constrain at present $R'^{-1}$ to be
above about 3--4~TeV~\cite{prec,well}. Nevertheless, LHC still has the chance of
observing the first Kaluza--Klein excitations of Standard Model particles
or, at least, of setting bounds on $R'^{-1}$ of more than 6~TeV~\cite{well,qui}.

If indeed quantum gravity sets in at the electroweak scale, future
collider experiments will directly test the structure of its
unknown dynamics. For instance, if string theory becomes relevant at 
$M_D$~\cite{clyk}, experiments could observe Regge recurrences with
higher masses and spins. It is certain that, whatever the underlying
weak-scale
quantum gravity theory may be, collider experiments in the TeV range will
be quite exciting.

\section{Conclusions}

We are now entering a phase in which searches for new physics are 
becoming the main experimental goal. The community in theoretical 
physics beyond the Standard Model is therefore facing 
a special responsibility. I believe that we are responding
to this challenge, since in the last few years numerous
new theoretical ideas have arisen to question some of the traditional
beyond-the-Standard-Model assumptions. It is too early to make definite
assessments, but it is very plausible to believe that some of these ideas
may lead to a profound revision of our views on the underlying high-energy
theory.

In this talk, I first made a few theoretical comments on  neutrino
oscillation data, the first direct indication of physics beyond the
Standard Model. Then, I turned to discussing supersymmetry and showed how 
recent research has focused on the problem of the ultraviolet sensitivity
of the soft terms. Solutions to this problem yield control over
flavour violations and calculability of the supersymmetric mass spectrum.
Finally, I discussed some recent developments in theories with extra
dimensions, aiming at bringing the gravitational scale down to the TeV
region. These proposals require a complete rethinking of the high-energy
behaviour in theories beyond the Standard Model. Therefore they have
deep physical and cosmological implications, beside the more sociological
implication of bringing closer together formal research and phenomenology.
If these theories are true, collider experiments
will observe a great deal of 
surprises above the TeV.


\bigskip
I am grateful to R.~Barbieri, G.~Buchalla, 
S.~Dimopoulos, F.~Feruglio, T.~Gherghetta,
M.~Mangano, J.~March-Russell, A.~Masiero, A.~Nelson, M.~Peskin, 
A.~Pomarol, R.~Rattazzi, A.~Riotto, A.~Strumia, and
J.~Wells for useful discussions
during the preparation of this talk.


\begin{thebibliography}{99}

\bibitem{superk} 
Y. Fukuda {\it et al.} (Super-Kamiokande Coll.), 
hep-ex/9805006, hep-ex/9805021, and  hep-ex/9807003. 

\bibitem{atmos}
S. Hatakeyama {\it et al.} 
(Kamiokande Coll.), hep-ex/9806038;
M. Ambrosio {\it et al.} (MACRO Coll.), hep-ex/9807005; 
M. Spurio (for the MACRO Coll.),
hep-ex/9808001;
W.W.M. Allison {\it et al.} (Soudan-2 Coll.),
hep-ex/9901024. 

\bibitem{tmann} T. Mann, these Proceedings.

\bibitem{neutr} For a review and list of references, see 
G.~Altarelli and F.~Feruglio,
hep-ph/9905536.

\bibitem{alta}
G.~Altarelli and F.~Feruglio,
Phys.\ Lett.\ {\bf B451}, 388 (1999).

\bibitem{alta2}
S.F.~King,
Phys.\ Lett.\ {\bf B439}, 350 (1998);
S.~Davidson and S.F.~King,
Phys.\ Lett.\ {\bf B445}, 191 (1998);
G.~Altarelli, F.~Feruglio and I.~Masina,
hep-ph/9907532.



\bibitem{lepto}
M.~Fukugita and T.~Yanagida,
Phys.\ Lett.\ {\bf B174}, 45 (1986).

\bibitem{franc} V. Ruhlmann-Kleider, these Proceedings.

\bibitem{natural1} 
J.~Ellis, K.~Enqvist, D.V.~Nanopoulos and F.~Zwirner,
Mod.\ Phys.\ Lett.\ {\bf A1}, 57 (1986).

\bibitem{natural2}
R.~Barbieri and G.F.~Giudice,
Nucl.\ Phys.\ {\bf B306}, 63 (1988).


\bibitem{natural3}
G.W.~Anderson and D.J.~Castano,
Phys.\ Lett.\ {\bf B347}, 300 (1995).



\bibitem{romans} 
L.~Giusti, A.~Romanino and A.~Strumia,
Nucl.\ Phys.\ {\bf B550}, 3 (1999).

\bibitem{pol}
P.H.~Chankowski, J.~Ellis, M.~Olechowski and S.~Pokorski,
Nucl.\ Phys.\ {\bf B544}, 39 (1999).

\bibitem{feng}
J.L.~Feng, K.T.~Matchev and T.~Moroi,
hep-ph/9909334.

\bibitem{wri}
D.~Wright,
hep-ph/9801449;
G.L.~Kane and S.F.~King,
Phys.\ Lett.\ {\bf B451}, 113 (1999);
M.~Bastero-Gil, G.L.~Kane and S.F.~King,
hep-ph/9910506.

\bibitem{grav}
A.H.~Chamseddine, R.~Arnowitt and P.~Nath,
Phys.\ Rev.\ Lett.\ {\bf 49}, 970 (1982);
R.~Barbieri, S.~Ferrara and C.A.~Savoy,
Phys.\ Lett.\ {\bf B119}, 343 (1982).



\bibitem{kos}
L.J.~Hall, V.A.~Kostelecky and S.~Raby,
Nucl.\ Phys.\ {\bf B267}, 415 (1986).



\bibitem{hall}
R.~Barbieri and L.J.~Hall,
Phys.\ Lett.\ {\bf B338}, 212 (1994);
R.~Barbieri, L.~Hall and A.~Strumia,
Nucl.\ Phys.\ {\bf B445}, 219 (1995).


\bibitem{gaug}
M.~Dine and A.E.~Nelson,
Phys.\ Rev.\ {\bf D48}, 1277 (1993);
M.~Dine, A.E.~Nelson and Y.~Shirman,
Phys.\ Rev.\ {\bf D51}, 1362 (1995);
M.~Dine, A.E.~Nelson, Y.~Nir and Y.~Shirman,
Phys.\ Rev.\ {\bf D53}, 2658 (1996);
G.F.~Giudice and R.~Rattazzi,
hep-ph/9801271.

\bibitem{lis}
L.~Randall and R.~Sundrum,
hep-th/9810155.

\bibitem{noi}
G.F.~Giudice, M.A.~Luty, H.~Murayama and R.~Rattazzi,
JHEP {\bf 12}, 027 (1998).

\bibitem{ricc}
A.~Pomarol and R.~Rattazzi,
JHEP {\bf 05}, 013 (1999).

\bibitem{alt}
Z.~Chacko, M.A.~Luty, I.~Maksymyk and E.~Ponton,
hep-ph/9905390;
E.~Katz, Y.~Shadmi and Y.~Shirman,
JHEP {\bf 08}, 015 (1999).

\bibitem{lis2}
J.L.~Feng, T.~Moroi, L.~Randall, M.~Strassler and S.~Su,
Phys.\ Rev.\ Lett.\ {\bf 83}, 1731 (1999).

\bibitem{noi2}
T.~Gherghetta, G.F.~Giudice and J.D.~Wells,
hep-ph/9904378.


\bibitem{ktev} A. Alavi-Harati {\it et al.} (KTeV Coll.),
Phys.\ Rev.\ Lett.\ {\bf 83}, 22 (1999);
E.~Blucher, these Proceedings.

\bibitem{na48} V. Fanti {\it et al.} (NA48 Coll.),
hep-ex/9909022;
G.~Barr, these Proceedings.

\bibitem{dag} G. D'Agostini, hep-ex/9910036.

\bibitem{eps}
A.~Buras, M.~Jamin and M.E.~Lautenbacher,
Nucl.\ Phys.\ {\bf B408}, 209 (1993);
M.~Ciuchini, E.~Franco, G.~Martinelli and L.~Reina,
Nucl.\ Phys.\ {\bf B415}, 403 (1994);
S.~Bosch, A.J.~Buras, M.~Gorbahn, S.~Jager, M.~Jamin, 
M.E.~Lautenbacher and L.~Silvestrini,
hep-ph/9904408;
M.~Ciuchini, E.~Franco, L.~Giusti, V.~Lubicz and G.~Martinelli,
hep-ph/9910237;
M.~Jamin,
hep-ph/9911390.

\bibitem{gabr}
E.~Gabrielli and G.F.~Giudice,
Nucl.\ Phys.\ {\bf B433}, 3 (1995).

\bibitem{mura}
A.~Masiero and H.~Murayama,
Phys.\ Rev.\ Lett.\ {\bf 83}, 907 (1999).

\bibitem{epss}
K.S.~Babu, B.~Dutta and R.N.~Mohapatra, hep-ph/9905464;
S.~Khalil and T.~Kobayashi, hep-ph/9906373;
R.~Barbieri, R.~Contino and A.~Strumia, hep-ph/9908255;
G.~Eyal, A.~Masiero, Y.~Nir and L.~Silvestrini,
JHEP {\bf 11}, 032 (1999).

\bibitem{math}
A.L.~Kagan and M.~Neubert,
hep-ph/9908404.

\bibitem{dort} T.~Hambye, G.O.~Kohler, E.A.~Paschos and P.H.~Soldan,
hep-ph/9906434;
J.~Bijnens and J.~Prades,
JHEP {\bf 01}, 023 (1999).

\bibitem{trie} S.~Bertolini, M.~Fabbrichesi and J.O.~Eeg,
hep-ph/9802405.

\bibitem{germ}
S.~Gardner and G.~Valencia,
hep-ph/9909202.

\bibitem{pich}
E.~Pallante and A.~Pich,
hep-ph/9911233.


\bibitem{add}
N.~Arkani-Hamed, S.~Dimopoulos and G.~Dvali,
Phys.\ Lett.\ {\bf B429}, 263 (1998) and
Phys.\ Rev.\ {\bf D59}, 086004 (1999).

\bibitem{brane}J. Polchinski, hep-th/9611050;
C.P. Bachas, hep-th/9806199.

\bibitem{mode}
P.~Horava and E.~Witten,
Nucl.\ Phys.\ {\bf B460}, 506 (1996) and Nucl.\ Phys.\ {\bf B475}, 94 (1996);
I.~Antoniadis, N.~Arkani-Hamed, S.~Dimopoulos and G.~Dvali,
Phys.\ Lett.\ {\bf B436}, 257 (1998);
Z.~Kakushadze and S.H.~Tye,
Nucl.\ Phys.\ {\bf B548}, 180 (1999).


 
\bibitem{stab}R.~Sundrum,
Phys.\ Rev.\ {\bf D59}, 085010 (1999);
N.~Arkani-Hamed, S.~Dimopoulos and J.~March-Russell,
hep-th/9809124.

\bibitem{rs1}
L.~Randall and R.~Sundrum,
Phys.\ Rev.\ Lett.\ {\bf 83}, 3370 (1999).

\bibitem{rl}
J.~Lykken and L.~Randall,
hep-th/9908076.

\bibitem{rs2}
L.~Randall and R.~Sundrum,
hep-th/9906064.

\bibitem{cohe}
A.G.~Cohen and D.B.~Kaplan,
hep-th/9910132.

\bibitem{flav}
N.~Arkani-Hamed and S.~Dimopoulos,
hep-ph/9811353;
Z.~Berezhiani and G.~Dvali,
Phys.\ Lett.\ {\bf B450}, 24 (1999).

\bibitem{yukk}
N.~Arkani-Hamed, L.~Hall, D.~Smith and N.~Weiner,
hep-ph/9909326.

\bibitem{yukkk}
N.~Arkani-Hamed and M.~Schmaltz,
hep-ph/9903417.

\bibitem{neutrhd}
K.R.~Dienes, E.~Dudas and T.~Gherghetta,
hep-ph/9811428;
N.~Arkani-Hamed, S.~Dimopoulos, G.~Dvali and J.~March-Russell,
hep-ph/9811448;
A.E.~Faraggi and M.~Pospelov,
Phys.\ Lett.\ {\bf B458}, 237 (1999).

\bibitem{gauc}
K.R.~Dienes, E.~Dudas and T.~Gherghetta,
Phys.\ Lett.\ {\bf B436}, 55 (1998) and
Nucl.\ Phys.\ {\bf B537}, 47 (1999).


\bibitem{gauc2}
N.~Arkani-Hamed, S.~Dimopoulos and J.~March-Russell,
hep-th/9908146.

\bibitem{dav}K.~Benakli and S.~Davidson,
Phys.\ Rev.\ {\bf D60}, 025004 (1999).

\bibitem{culle}
S.~Cullen and M.~Perelstein,
Phys.\ Rev.\ Lett.\ {\bf 83}, 268 (1999).

\bibitem{halgr} L.J.~Hall and D.~Smith,
Phys.\ Rev.\ {\bf D60}, 085008 (1999).

\bibitem{pirl}
J.C.~Long, H.W.~Chan and J.C.~Price,
Nucl.\ Phys.\ {\bf B539}, 23 (1999).

\bibitem{sav}
S.~Dimopoulos and G.F.~Giudice,
Phys.\ Lett.\ {\bf B379}, 105 (1996).

\bibitem{giu}
G.F.~Giudice, R.~Rattazzi and J.D.~Wells,
Nucl.\ Phys.\ {\bf B544}, 3 (1999).

\bibitem{pesk}
E.A.~Mirabelli, M.~Perelstein and M.E.~Peskin,
Phys.\ Rev.\ Lett.\ {\bf 82}, 2236 (1999);
T.~Han, J.D.~Lykken and R.~Zhang,
Phys.\ Rev.\ {\bf D59}, 105006 (1999).

\bibitem{hew}
J.L.~Hewett, Phys.\ Rev.\ Lett.\ {\bf 82}, 4765 (1999).

\bibitem{anto}
I.~Antoniadis,
Phys.\ Lett.\ {\bf B246}, 377 (1990).

\bibitem{prec}
P. Nath and M. Yamaguchi, hep-ph/9902323; M. Masip and A. Pomarol,
hep-ph/9902467; W.J. Marciano, hep-ph/9903451;
A.~Strumia,
hep-ph/9906266.

\bibitem{well} T.G. Rizzo and J.D. Wells, hep-ph/9906234.

\bibitem{qui}
I.~Antoniadis, K.~Benakli and M.~Quiros,
Phys.\ Lett.\ {\bf B331}, 313 (1994) and
hep-ph/9905311.

\bibitem{clyk}
J.D.~Lykken,
Phys.\ Rev.\ {\bf D54}, 3693 (1996).


\end{thebibliography}
\end{document}